\documentclass[twoside]{article}
\usepackage[margin=1in]{geometry}
\usepackage{authblk}

\usepackage{graphicx} 
\usepackage[colorlinks=true, allcolors=black]{hyperref}
\usepackage{amsmath}
\usepackage{amssymb}
\usepackage{fancyhdr}
\usepackage{multirow}

\title{\textbf{Dose-PlanNet: Physics Based Radiotherapy Dose Prediction with Deep Learning}}

\author[1,2]{Ankit Bhattacharjee}
\author[2]{Sougata Maity}
\author[2]{Santam Chakraborty}
\author[2]{Indranil Mallick}

\affil[1]{Department of Mechanical Engineering, Indian Institute of Technology Kharagpur}
\affil[2]{Department of Radiation Oncology, Tata Medical Center Kolkata}

\date{\today}

\begin{document}

\maketitle

\begin{abstract}
Automating prostate radiotherapy treatment planning is dosimetrically complex, particularly for extreme hypofractionated regimens. In this study, we introduce Dose-PlanNet, a physics-guided 3D deep learning architecture designed to predict dose distributions. This model's performance was evaluated on a cohort of patients treated in a prospective trial where two different dose fractionation regimens were employed. Dose-PlanNet achieved comparable target coverage ($D_{95}$), though statistical analysis revealed a marginal reduction in target homogeneity ($p<0.001$) offset. However the model achieved statistically significant improvements in high-dose organ-at-risk sparing ($p<0.001$). When evaluated against strict Prospective Randomized protocol volumetric constraints, automated plans met prespecified clinical acceptance criteria in 11 out of 14 Moderate Hypofraction Arm plans and 9 out of 12 Stereotactic Body Radiation Therapy Arm plans. This pipeline demonstrates that physics-informed deep learning can accelerate radiotherapy workflows while safely maintaining the stringent dosimetric quality required for high-precision clinical deployment.
\end{abstract}

\section{Introduction}
The standard of care for high-risk localized prostate cancer is Intensity-Modulated Radiation Therapy (IMRT) \cite{wu2009simultaneous, dogan2003imrt}. When irradiation of the pelvic nodes is performed, a simultaneous integrated boost (SIB) approach is used. SIB regimens require the simultaneous delivery of an escalated prescription dose to the primary planning target volume (PTV) and a lower dose to the elective pelvic lymph nodes. These plans require complex spatial dose gradients \cite{wu2009simultaneous}. This multi-target dosimetric objective must be achieved while strictly adhering to mandatory maximum and dose-volume constraints for nearby organs at risk (OARs), specifically the anorectum and bladder, to prevent severe gastrointestinal and genitourinary toxicities \cite{michalski2010toxicity}. Consequently, manual inverse treatment planning remains a highly iterative, resource-intensive workflow that is heavily dependent on the individual planner's experience. This invariably leads to significant inter-planner variability and sub-optimal operational efficiency in high-volume clinical settings \cite{nelms2012variation}.

Recent advancements in deep learning, particularly 3D U-Net architectures, have demonstrated significant potential in automating volumetric dose prediction \cite{nguyen20193dUnet, kearney2018doseNet}. To mitigate the resource-intensive nature of manual optimization, recent clinical workflows have successfully leveraged knowledge-based planning (KBP) frameworks to fully automate treatment plan generation across diverse anatomical sites \cite{chung2025knowledge}. However, conventional frameworks predominantly approach dose prediction as a standard image-to-image translation task, relying on Mean Squared Error (MSE) or Mean Absolute Error (MAE) as the primary optimization metric \cite{babier2018knowledge}. This creates a fundamental disconnect between mathematical optimization and clinical reality. MSE intrinsically minimizes the global average voxel-wise error, meaning it treats a 1 Gy deviation in the geometric center of the target volume with the exact same mathematical weight as a 1 Gy deviation spilling across the boundary into the rectal wall. Clinical plan acceptability, however, is not determined by global volumetric averages; it is strictly governed by localized boundary constraints, absolute maximum dose ceilings, and specific dose-volume histogram (DVH) thresholds. Consequently, MSE-driven models frequently generate dose distributions that achieve a low global error but fail clinical evaluation due to point-dose violations in serial OARs or unacceptable cold spots at the target periphery \cite{mcintosh2017voxel}. To bridge the gap between theoretical volumetric approximation and clinical deployability, neural network architectures must move beyond generic pixel-wise loss. They require physics-guided optimization engines capable of spatial awareness, prioritizing critical tissue boundaries, and mimicking the non-linear dosimetric compromises executed by expert human planners.

To overcome the limitations of generic voxel-wise objectives, a treatment planning framework must transition from statistical image translation to mathematically enforced, physics-guided optimization \cite{ge2019dvh}. Achieving true clinical usability dictates that an artificial intelligence must ``see'' anatomical boundaries rather than merely mapping isolated voxel intensities. This requires explicit spatial awareness of the local anatomy, allowing the network to calculate the exact proximity of high-dose clouds to abutting serial structures \cite{chen2019incorporating}. It must understand the fundamental radiation dose absorption principles like the inverse square law which drives these gradients such that absurd dose gradients are not generated.  Furthermore, the optimization engine must deploy a multi-objective penalty system that inherently understands the hierarchy of clinical constraints. It must mathematically differentiate between a soft penalty for optional dose spillage into generic healthy tissue and a severe, overriding penalty for a mandatory constraint violation within the rectal wall or bladder \cite{mahmood2018automated}. By encoding these strict dosimetric rules directly into the gradient descent space, the network is forced to learn the non-linear spatial compromises required for safe, physician-approvable clinical planning.

In this work, we propose \textbf{Dose-PlanNet}, a physics-guided deep learning framework specifically engineered to generate clinically acceptable IMRT plans for prostate Simultaneous Integrated Boost (SIB) protocols. The novelty of Dose-PlanNet is anchored in three primary methodological contributions. First, we utilize discrete PTV mapping, governed by a Painter's Algorithm, to explicitly inject varying prescription doses directly into the input tensor, definitively resolving SIB dose hierarchies. Second, we equip the network with absolute spatial awareness by integrating Signed Distance Maps (SDMs) for abutting serial organs alongside a Beam's Eye View (BEV) frustum mask; this combination allows the architecture to mathematically distinguish in-beam dose corridors from physically impossible out-of-beam scatter. Finally, we replace standard voxel-wise metrics with a custom, multi-objective physics-guided loss engine. This engine features differentiable Dose-Volume Histogram (DVH) estimations via steep sigmoid approximations, deploying tuned penalties to strictly enforce global dose ceilings, exact V-type OAR constraints, and absolute PTV coverage thresholds. Evaluated on a clinical prostate SIB cohort, Dose-PlanNet successfully generated steep dose gradients to yield physician-approvable dose distributions. The framework achieved good OAR sparing while simultaneously satisfying mandatory PTV coverage constraints for the vast majority of patients, demonstrating robust clinical utility over standard translation-based AI approaches.

\section{Related Work}
The automation of radiotherapy dose prediction has evolved significantly over the past decade, transitioning from generalized medical image processing to highly specialized dosimetric optimization. This section reviews the progression of deep learning frameworks in this domain, highlighting the persistent clinical limitations that necessitate a physics-guided approach.

\subsection{Volumetric Dose Prediction as Image Translation}
The foundational era of deep learning-based dosimetry primarily conceptualized dose prediction as a standard image-to-image translation task. Early frameworks heavily relied on 3D U-Net architectures \cite{nguyen20193dUnet} and Fully Convolutional Neural Networks (FCNs), such as DoseNet \cite{kearney2018doseNet}, to map anatomical computed tomography (CT) and contour data directly to dose distributions. Subsequently, Generative Adversarial Networks (GANs) were introduced to further refine the global texture and statistical distribution of the predicted volumetric dose clouds \cite{babier2018knowledge}. 

\subsection{Anatomical and Spatial Embeddings}
To address the spatial blindness of pure image translation, subsequent research introduced explicit anatomical embeddings. Frameworks began incorporating Signed Distance Maps (SDMs) and distance-to-target coordinates directly into the input tensor to force network awareness of critical tissue boundaries \cite{chen2019incorporating}. While providing this spatial context significantly improved the geometric conformity of the predicted dose clouds, it failed to resolve the fundamental optimization bottleneck. Because these architectures still minimized standard MSE or MAE loss, they remained mathematically incapable of prioritizing strict clinical boundary constraints over bulk tissue averages. Consequently, despite ``seeing" the organs, these models remained highly susceptible to localized but clinically fatal dosimetric violations \cite{ge2019dvh}.

\subsection{Physics-Guided and Constraint-Aware Optimization}
To overcome the inherent limitations of MSE, recent state-of-the-art frameworks have begun encoding clinical dosimetry rules directly into the network's optimization trajectory. Notably, architectures such as nnDoseNet have demonstrated the utility of embedding differentiable Dose-Volume Histogram (DVH) metrics into the loss space, forcing the network to optimize for clinical acceptability rather than pure voxel matching \cite{nndosenet_preprint}. Concurrently, recent experimental approaches in physics-guided radiotherapy have introduced dynamic penalty engines to mathematically enforce physical dose delivery limits \cite{physics_guided_preprint}. 

While these constraint-aware engines represent a critical methodological leap, their evaluation has predominantly been restricted to single-target prescriptions, leaving the multi-objective complexities of Simultaneous Integrated Boost (SIB) protocols largely unaddressed. Furthermore, without steep, dynamically weighted penalty gradients, these loss engines risk severe target underdosing, forcing the network into sub-optimal mathematical tradeoffs between target saturation and strict OAR sparing.

\subsection{The Unsolved SIB and Beam Geometry Gap}
Despite advancements in physics-guided optimization, existing frameworks remain fundamentally ill-equipped for the spatial complexities of Simultaneous Integrated Boost (SIB) protocols. Many conventional frameworks rely on implicit spatial approximations rather than dedicated mathematical mechanisms to definitively resolve abutting or overlapping target volumes that require distinct prescription doses. Furthermore, they almost universally ignore the physical limitations of beam geometry. By failing to restrict optimization space to deliverable beam trajectories, these architectures frequently artificially minimize their loss functions by depositing low-dose scatter into physically impossible out-of-beam regions. 

Dose-PlanNet directly targets these limitations. By implementing a Painter's Algorithm for discrete PTV mapping \cite{newell1972solution}, the framework explicitly enforces strict SIB dose hierarchies to prevent target overlap ambiguity. Concurrently, the integration of a Beam's Eye View (BEV) frustum mask into the multi-objective penalty engine mathematically prohibits non-physical out-of-beam dose spillage in an IMRT plan. This methodology bridges the remaining gap between theoretical physics-guided prediction and verifiable clinical deployability.

\section{Methodology}
The development of Dose-PlanNet requires a multi-stage pipeline that transitions from raw clinical data curation to mathematically constrained neural network optimization. This section details the spatial data embeddings, the volumetric network architecture, and the multi-objective physics-guided loss engine that governs the framework.

\begin{figure*}[ht!]
    \centering
    \includegraphics[width=\textwidth]{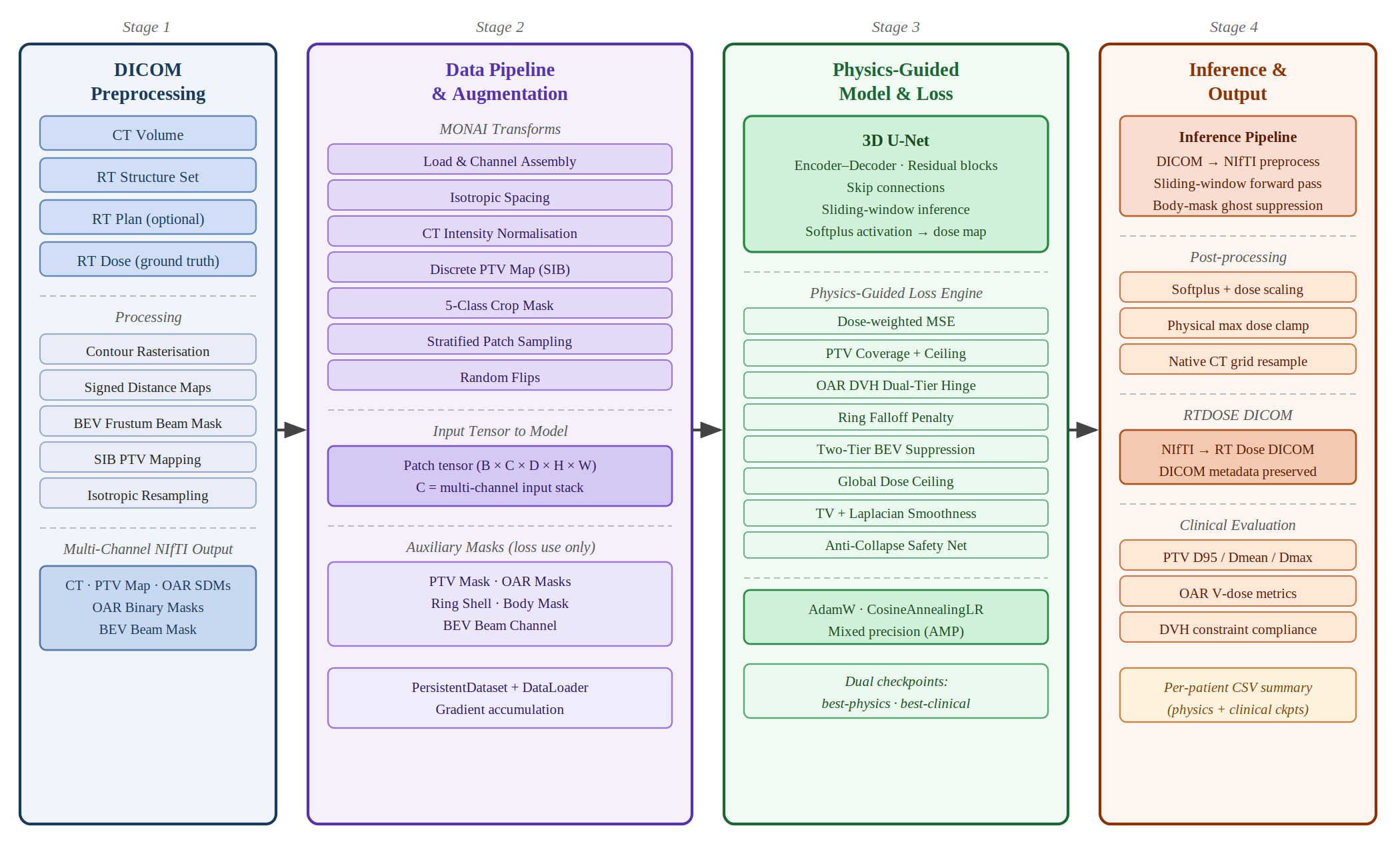}
    \caption{\textbf{End-to-End Architecture of the Dose-PlanNet Pipeline.} Raw clinical DICOM files are isotropically resampled and converted into physics-aware spatial embeddings, including Signed Distance Maps (SDMs), Beam's Eye View (BEV) frustums, and a discrete Simultaneous Integrated Boost (SIB) map via Painter's algorithm. The 7-channel tensor feeds a 3D U-Net trained via a multi-objective physics loss engine, where clinical constraints are progressively introduced during early training. The pipeline yields dual checkpoints (Physics-Optimized and Clinically-Optimized), which are utilized during sliding-window inference. Crucially, the final step mathematically reverses the affine resampling operations, restoring the predicted dose matrix to the native CT geometry for immediate Treatment Planning System (TPS) integration.}
    \label{fig:pipeline}
\end{figure*}

\subsection{Data Representation and Preprocessing}
The translation of raw, unorganized clinical DICOM data into a structured tensor format suitable for deep learning optimization is a critical bottleneck in volumetric dosimetry. To address this, we developed a modular preprocessing pipeline that automates the alignment and digitization of patient anatomy, treatment geometries, and prescribed dose objectives.

\subsubsection{Anatomical Digitization and Distance Mapping}
The foundational coordinate system for each patient is established by loading the computed tomography (CT) scan and performing isotropic voxel resampling to a standardized resolution of $1.27 \times 1.27 \times 2.5 \text{ mm}^{3}$. To generate the external body contour,a threshold based segmentation (HU -350 or higher) is done.

To ensure the network explicitly recognizes the physical boundaries of adjacent anatomical structures, physician-defined contours (RTStruct) are digitized into voxelized binary masks. Rather than relying solely on these binary representations, the pipeline computes Signed Distance Maps (SDMs) for a configurable list of abutting serial Organs at Risk (OARs), specifically the bladder and Anorectum in this work. The SDM calculates the shortest physical distance from every spatial coordinate $\vec{r}$ to the nearest organ boundary:
\begin{equation}
SDM(\vec{r}) = \begin{cases} 
0, & \text{if } \vec{r} \in \mathcal{S} \\
d(\vec{r}, \partial \mathcal{S}), & \text{if } \vec{r} \notin \mathcal{S} 
\end{cases}
\end{equation}
where $\mathcal{S}$ represents the specific structure volume and $d(\vec{r}, \partial \mathcal{S})$ is the Euclidean distance to the structure boundary. This continuous spatial embedding forces the optimization engine to recognize the strict requirement for steep dose falloff immediately adjacent to healthy tissue.

\subsubsection{Discrete SIB Mapping via Painter's Algorithm}
A fundamental challenge in automating Simultaneous Integrated Boost (SIB) protocols is the ambiguity of target intersections, where elective target volumes (with a lower dose prescription) overlap with the primary target where a higher dose is to be delivered. To resolve this spatial conflict, the pipeline executes a discrete PTV mapping strategy governed by a Painter's Algorithm \cite{newell1972solution}.

Individual target volumes are extracted and sorted in ascending order based on their prescription dose (e.g., $PTV_{44}$ followed by $PTV_{62}$). The algorithm sequentially paints these volumes into a single discrete tensor channel, assigning the exact numerical prescription dose (in Gray) as the voxel intensity. Because higher-dose targets are rasterized last, they overwrite the overlapping low-dose regions. This explicitly preserves the dosimetric hierarchy, providing the network with an unambiguous ground-truth target landscape.

\subsubsection{Beam's Eye View (BEV) Frustum Construction}
To physically constrain the network from depositing radiation in geometrically impossible regions, the pipeline extracts the fixed gantry angles from the clinical RTPlan file to construct a Beam's Eye View (BEV) frustum mask. 

For a given gantry angle $\theta$, the beam direction vector is defined as $\hat{\Omega}_{\theta} = [\sin\theta, -\cos\theta, 0]^T$. The source position $\vec{S}$ is calculated using the established Source-to-Axis Distance ($SAD = 1000\text{ mm}$) relative to the PTV isocenter. The pipeline performs geometric ray-tracing to determine if a spatial voxel $\vec{r}$ lies within the active beam corridor. A voxel is classified as ``In-Beam" if its perpendicular distance to the central axis is bounded by the projected target radius and an isotropic penumbra margin:
\begin{equation}
BEV(\vec{r}) = \begin{cases} 
1, & \text{if } (\vec{r}-\vec{S}) \cdot \hat{\Omega}_{\theta} > 0 \text{ and } d_{\perp} < d_{\parallel} \tan\phi + p_{\text{margin}} \\
0, & \text{otherwise}.
\end{cases}
\end{equation}
where $d_{\perp}$ and $d_{\parallel}$ are the perpendicular and parallel distances relative to the source, $\phi$ is the half-angle subtended by the target, and $p_{margin}$ is the $7.0\text{ mm}$ physical penumbra. This binary mask is subsequently passed to the loss engine to violently penalize non-physical scatter.

\subsection{Network Architecture}
The core dose prediction engine utilizes a 3D U-Net architecture to map the prepared spatial tensors to a continuous volumetric dose distribution. The network ingests a 7-channel input tensor: [CT, Discrete PTVs, Bladder SDM, Anorectum SDM, Body Mask, Penile Bulb Mask, BEV Frustum].

The architecture employs an encoder-decoder topology with four resolution levels (channel dimensions of 16, 32, 64, 128) and skip connections to preserve high-frequency spatial boundaries. To ensure numerical stability and physical realism, a Softplus activation function, $f(x) = \ln(1+e^x)$, is applied to the terminal layer. This mathematically guarantees a strict non-negative physical dose floor ($D_{\text{pred}} \ge 0$) while avoiding the vanishing gradient problem associated with hard clipping functions in high-dose regions.

\paragraph{Auxiliary Spatial Masks for Loss Optimization.}
While the primary 7-channel tensor provides the necessary spatial context for the network's forward pass, several critical structures are routed directly to the loss engine to exclusively govern backpropagation penalties. These auxiliary inputs include the binary masks for the bowel bag and the bilateral femoral heads, which are used strictly to enforce dose-volume and maximum dose constraints without cluttering the network's anatomical input space. 

Furthermore, to mathematically enforce the steep dose gradients required immediately outside the target volumes, the preprocessing pipeline dynamically computes a 3D geometric falloff ring. This is achieved by applying a 3D morphological max-pooling dilation to the discrete PTV mask. Given the anisotropic voxel spacing of $1.27 \times 1.27 \times 2.5 \text{ mm}^{3}$, a non-uniform kernel of $5 \times 9 \times 9$ voxels is utilized to generate a precise 5.0 mm isotropic expansion. The original PTV is then subtracted from this dilated volume, yielding a hollow, binary geometric shell:
\begin{equation}
\text{Ring}_{5mm} = \text{MaxPool3D}(PTV, k_{5\times9\times9}) - PTV.
\end{equation}
This isolated 5 mm boundary corridor is passed exclusively to the multi-objective loss engine to heavily penalize dose spillage immediately adjacent to the prescription boundary.

\subsection{The Physics-Guided Loss Engine}
The fundamental limitation of generic image translation networks is their reliance on unweighted Mean Squared Error (MSE), which mathematically values low-dose background approximations equally against critical high-dose target conformity. To force the network to prioritize clinical acceptability, Dose-PlanNet optimizes a custom, multi-objective loss function that integrates the spatial, geometric, and dosimetric penalties defined throughout this section:
\begin{equation}
\label{loss}
\begin{split}
\mathcal{L}_{total} &= \lambda_{mse}\mathcal{L}_{MSE} + \lambda_{ptv}\mathcal{L}_{PTV} + \lambda_{v}\mathcal{L}_{V\text{-}Type} \\
&\quad + \lambda_{ring}\mathcal{L}_{ring} + \lambda_{ac}\mathcal{L}_{anticollapse} + \lambda_{body}\mathcal{L}_{body} \\
&\quad + \lambda_{bg}\mathcal{L}_{BEV} + \lambda_{reg}\mathcal{L}_{Smooth},
\end{split}
\end{equation}
where the tuned $\lambda$ coefficients balance voxel-wise fidelity against strict clinical boundary constraints.

\begin{figure}[h!]
    \centering
    \includegraphics[width=\textwidth, height=15cm, keepaspectratio]{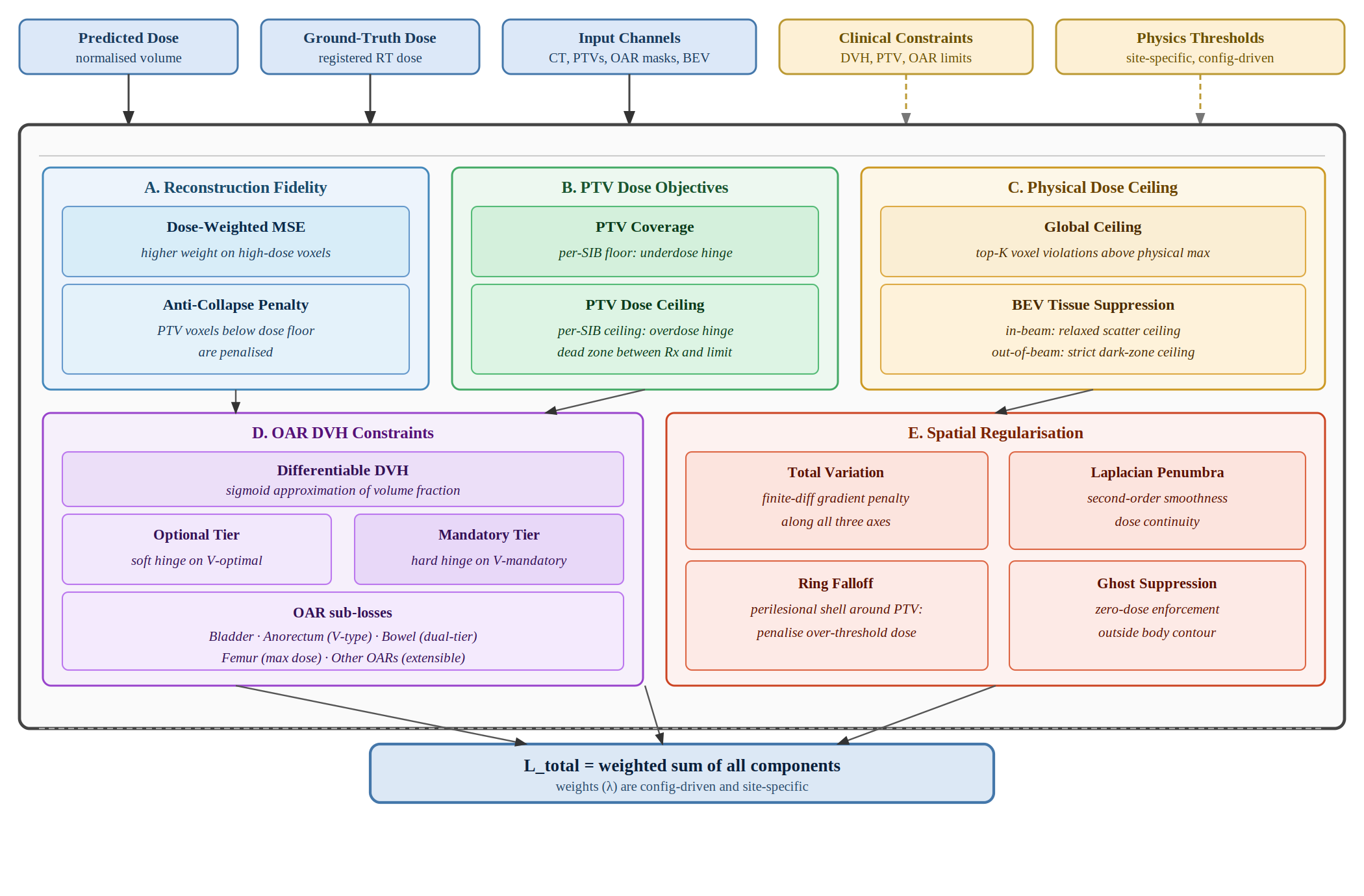}
    \caption{\textbf{Decomposition of the Physics-Guided Loss Engine.} The network optimizes a custom multi-objective loss function ($\mathcal{L}_{total}$) driven by complex spatial embeddings. Constraints are categorized into Absolute Dose penalties (D-Type, targeting specific dose thresholds like PTV coverage and anti-collapse), Volume-Histogram penalties (V-Type, utilizing a differentiable sigmoid estimator for OAR sparing), and Spatial/Regularization penalties (enforcing geometric falloff and scatter suppression). A curriculum scheduler dynamically ramps the magnitude of these physics penalties from zero over the initial 30 epochs to ensure early numerical stability.}
    \label{fig:loss_engine}
\end{figure}

\subsubsection{Dose-Weighted Mean Squared Error}
To prevent the optimizer from diluting its gradient updates across the vast volume of non-target background tissue, the baseline volumetric reconstruction loss is modified into a Dose-Weighted Mean Squared Error. A spatial weighting matrix, $W(\vec{r})$, is computed from the ground-truth dose to exponentially amplify the penalty in high-dose prescription regions:
\begin{equation}
W(\vec{r}) = 1 + \alpha \cdot \min\{D_{\text{true}}(\vec{r}), D_{\text{clamp}}\},
\end{equation}
where $\alpha = 9.0$ sets the maximum amplification scale and $D_{\text{clamp}} = 0.96$ prevents gradient explosion at the absolute hotspot maximum. The resulting weighted baseline loss is:
\begin{equation}
\mathcal{L}_{MSE} = \frac{1}{N} \sum_{\vec{r} \in \Omega} W(\vec{r}) \left[ D_{\text{pred}}(\vec{r}) - D_{\text{true}}(\vec{r}) \right]^2,
\end{equation}
where $\Omega$ denotes the entire three dimensional vector space in which the body exists. 

\subsubsection{D-Type (Absolute Dose) Penalties}
To enforce absolute target coverage and prevent hotspots, we implement D-Type penalty functions based on the prescription guidelines. For the Planning Target Volumes (PTVs), we utilize a one-sided hinge loss that penalizes underdosing relative to the prescription ($Rx$) and overdosing relative to the hard clinical ceiling ($D_{\text{ceil}}$):
\begin{equation}
\mathcal{L}_{PTV} = \frac{1}{N_{PTV}} \sum_{\vec{r} \in PTV} \left[ \lambda_{ptv} \text{ReLU}(D_{Rx} - D_{\text{pred}}(\vec{r}))^2 + \lambda_{ptv\_max} \text{ReLU}(D_{\text{pred}}(\vec{r}) - D_{\text{ceil}})^2 \right].
\end{equation}
Beyond the PTV, Dose-PlanNet enforces a global hard dose ceiling ($\text{global\_ceil\_gy}$) to suppress mathematically divergent hotspots across the entire body volume. To avoid gradient dilution in large-volume pelvic scans, we employ a Top-$K$ violation strategy, aggregating the penalty only from the $K=100$ most severe transgressions:
\begin{equation}
\mathcal{L}_{global\_ceil} = \frac{1}{K} \sum_{k=1}^{K} \text{ReLU}(D_{\text{pred}}^{(k)} - D_{\text{global\_ceil}})^2,
\end{equation}
where $D_{\text{pred}}^{(k)}$ are the $K$ largest positive violations of the ceiling threshold.

\subsubsection{Differentiable Dose-Volume Histogram (DVH) Estimator}
Clinical protocols mandate specific volumetric thresholds for serial organs, such as ensuring the anorectum volume receiving at least $60.4 \text{ Gy}$ ($V_{60.4\text{Gy}}$) remains below a strict percentage. Standard voxel counting relies on a non-differentiable Heaviside step function, which interrupts the computational graph during backpropagation. To resolve this, Dose-PlanNet approximates the step function using a steep sigmoid curve, facilitating continuous gradient flow. For a given anatomical structure $\mathcal{S}$, the predicted volume fraction exceeding a normalized dose threshold $D_{\text{ref}}$ is calculated as:
\begin{equation}
\label{eq:dvh}
V_{D_{\text{ref}}}(\mathcal{S}) \approx \frac{1}{N_{\mathcal{S}}} \sum_{\vec{r} \in \mathcal{S}} \sigma\left(k \cdot [D_{\text{pred}}(\vec{r}) - D_{\text{ref}}]\right),
\end{equation}
where $\sigma(x) = 1 / (1 + e^{-x})$, $N_{\mathcal{S}}$ is the total number of voxels within the structure mask, and $k = 50.0$ is the steepness multiplier. This formulation allows the optimizer to smoothly calculate volume fractions and apply differentiable hinge penalties, enabling the architecture to treat clinical $V$-type constraints as active, gradient-governed optimization objectives.

\subsubsection{V-Type (Volumetric) Dual-Tier Penalties}
Clinical protocols often define a dual-tier constraint system comprising an \textit{Optimal} volume threshold ($D_{\text{opt}}$) and a strictly enforced \textit{Mandatory} limit ($D_{\text{mand}}$). Using the differentiable DVH estimator defined in equation \ref{eq:dvh}, we formulate a two-stage penalty that forces the network to adhere to dosimetric objectives:
\begin{equation}
\mathcal{L}_{V-Type} = \sum_{c \in \mathcal{C}} \left[ \lambda_{opt} \text{ReLU}(V_{D_{\text{ref}}}(c) - D_{\text{opt},c})^2 + \lambda_{mand} \text{ReLU}(V_{D_{\text{ref}}}(c) - D_{\text{mand},c})^2 \right],
\end{equation}
where $V_{D_{\text{ref}}}(c)$ is the differentiable volume fraction at dose $D_{\text{ref}}$ for constraint $c$, and $\lambda_{mand} \gg \lambda_{opt}$ ensures that violating mandatory thresholds results in a disproportionately large gradient, forcing the optimizer to recover the plan within the clinical safe-zone.

\subsubsection{Spatial Scatter Suppression (BEV-Constrained)}
To eliminate physically impossible out-of-beam dose spillage, the BEV frustum mask, $M_{BEV} \in \{0, 1\}$, acts as a spatial filter for healthy tissue ($M_{Body} = 1$). We split the non-target tissue into In-Beam and Out-of-Beam corridors, applying a tiered suppression penalty:
\begin{equation}
\mathcal{L}_{BEV} = \frac{\lambda_{in}}{N_{in}} \sum_{\vec{r} \in \text{In-Beam}} \text{ReLU}(D_{\text{pred}}(\vec{r}) - D_{in}) + \frac{\lambda_{out}}{N_{out}} \sum_{\vec{r} \in \text{Out-of-Beam}} \text{ReLU}(D_{\text{pred}}(\vec{r}) - D_{out}),
\end{equation}
where $\lambda_{out} > \lambda_{in}$ and $D_{out} \ll D_{in}$. This enforces the physical reality that dose deposition outside the projected target geometry should be near-zero, effectively suppressing unmodeled scatter.

\subsubsection{Geometric Falloff, Anti-Collapse, and Ghost Suppression}
To mathematically enforce the steep dose gradients required immediately outside the prescription boundary, Dose-PlanNet evaluates the predicted dose against the 5.0 mm geometric falloff shell, denoted by $\mathcal{R}$, extracted during preprocessing. A one-sided quadratic penalty is applied to any voxel within this shell that exceeds the strict falloff threshold ($D_{\text{ring\_thresh}}$):
\begin{equation}
\mathcal{L}_{ring} = \frac{1}{N_{ring}} \sum_{\vec{r} \in \mathcal{R}} \text{ReLU}(D_{\text{pred}}(\vec{r}) - D_{\text{ring\_thresh}})^2.
\end{equation}

Concurrently, to safeguard against severe target drop-offs caused by competing OAR penalties, an anti-collapse safety net is embedded within the target volume. This aggressively penalizes any PTV voxel that falls below $50\%$ of the normalized prescription dose, preventing topological collapse in complex geometries:
\begin{equation}
\mathcal{L}_{anticollapse} = \left[ \frac{1}{N_{PTV}} \sum_{\vec{r} \in PTV} \text{ReLU}(0.50 - D_{\text{pred}}(\vec{r})) \right]^2.
\end{equation}

Furthermore, to prevent the network from mathematically displacing excess dose into the ambient air surrounding the patient (a common artifact in unconstrained 3D volumetric regression), an out-of-body dose suppression penalty is applied. This term strictly suppresses any non-zero dose predicted outside the binary body contour ($M_{\text{Body}}$):
\begin{equation}
\mathcal{L}_{body} = \frac{1}{N_{air}} \sum_{\vec{r} \notin \text{Body}} D_{\text{pred}}(\vec{r}).
\end{equation}

\subsubsection{Spatial Regularization}
To ensure the predicted dose distributions remain physically plausible, we incorporate spatial regularization terms that enforce smoothness and continuity across the volumetric grid. Total Variation (TV) is employed to penalize sharp, non-physical voxel intensity fluctuations:
\begin{equation}
\mathcal{L}_{smooth} = \sum_{\vec{r} \in \Omega} \left( \|\nabla_d D_{\text{pred}}(\vec{r})\|^2 + \|\nabla_h D_{\text{pred}}(\vec{r})\|^2 + \|\nabla_w D_{\text{pred}}(\vec{r})\|^2 \right),
\end{equation}
where $\nabla_d, \nabla_h, \nabla_w$ represent the first-order spatial gradients along the depth, height, and width axes, respectively. Furthermore, a Laplacian penalty is utilized to enforce second-order continuity, effectively suppressing high-frequency noise near the beam penumbra:
\begin{equation}
\mathcal{L}_{laplacian} = \sum_{\vec{r} \in \Omega} \left[ \nabla^2 D_{\text{pred}}(\vec{r}) \right]^2.
\end{equation}

\subsubsection{Total Objective Function}
The complete multi-objective loss function is the weighted summation of these complementary components, as formulated in Equation \ref{loss}.. During the curriculum training phase, all physics-based and geometric penalties (e.g., $\mathcal{L}_{PTV}$, $\mathcal{L}_{V\text{-}Type}$, $\mathcal{L}_{ring}$, $\mathcal{L}_{anticollapse}$, $\mathcal{L}_{body}$, and $\mathcal{L}_{BEV}$) are linearly ramped from zero to their target magnitudes over the initial $30$ epochs. This warmup strategy allows the 3D U-Net to first stabilize the global dose distribution via the constant, dose-weighted $\mathcal{L}_{MSE}$ before the more restrictive, non-linear clinical hinge penalties are introduced to finalize the dosimetric conformity and OAR sparing.

\section{Experimental Setup and Evaluation}
\label{exp}
To validate the clinical efficacy and generalization of the Dose-PlanNet architecture, this section details the retrospective patient cohorts across multiple fractionation schemes, the quantitative dosimetric criteria, and the blinded expert review methodology.

\subsection{Clinical Datasets and Patient Cohorts}
The dataset for this study comprises two distinct cohorts sourced from the Prostate Prospective Randomized trial conducted at Tata Medical Center, Kolkata. Active recruitment for the trial has concluded, establishing these as fixed, retrospective clinical datasets.

\begin{itemize}
    \item \textbf{Moderate Hypofraction Arm Cohort:} Comprises 70 clinical patients receiving standard fractionation. To rigorously evaluate generalization, this cohort was partitioned using an 80/20 train-test split, dedicating 56 patients to training and hyperparameter optimization, and strictly holding out 14 patients as an unseen testing set for final dosimetric inference.
    
    \item \textbf{Stereotactic Body Radiation Therapy Arm (SBRT) Cohort:} Comprises a separate set of 62 patients receiving Stereotactic Body Radiation Therapy (SBRT). This cohort was similarly partitioned into training and held-out validation sets to assess the network's capacity to map hypofractionated dose distributions and ultra-steep spatial gradients.
\end{itemize}

\subsection{Quantitative Evaluation Metrics}
The predicted dose distributions were mathematically evaluated against the strict protocol defined constraints \cite{murthy2020prime}. Because the cohorts utilize fundamentally different fractionation schemes, they were evaluated against separate dosimetric objectives:

\begin{itemize}
    \item \textbf{Moderate Hypofraction Arm (62.0 Gy Prescription):} The primary target metrics required a dosimetric coverage of $D_{95} \ge 95\%$ and strict adherence to an absolute maximum dose limit of 66.34 Gy. Evaluations for Organs at Risk (OARs) focused on specific volumetric thresholds, primarily the Bladder and Anorectum $V_{60.4\text{Gy}}$ and $V_{38.0\text{Gy}}$ constraints, alongside the maximum dose to the bilateral femoral heads ($D_{max} \le 40.0 \text{ Gy}$).
    \item \textbf{Stereotactic Body Radiation Therapy Arm (36.25 Gy Prescription):} The primary target metrics required a dosimetric coverage of $D_{95} \ge 95\%$ and strict adherence to an absolute maximum dose limit of 38.8 Gy. Evaluations for Organs at Risk (OARs) focused on specific volumetric thresholds, primarily the Bladder and Anorectum $V_{35.0\text{Gy}}$ and $V_{31.5\text{Gy}}$ constraints, alongside the volumetric dose limit to the bilateral femoral heads ($V_{14.0\text{Gy}} < 5\%$).
\end{itemize}

\subsection{Evaluated Baselines}
To rigorously validate the necessity of the physics-guided hinge penalties integrated into the Dose-PlanNet loss architecture, an ablation study was conducted using an unconstrained baseline configuration. This model was trained exclusively utilizing Mean Squared Error (MSE) loss, stripping away the other regularization terms and losses. This baseline was evaluated specifically on the Prostate Prospective Randomized Stereotactic Body Radiation Therapy Arm (SBRT) cohort to determine if a standard deep learning architecture could autonomously map hypofractionated dose distributions and navigate the ultra-steep spatial gradients required for clinical safety without explicit physics-informed guidance.

\subsection{Model Configurations and Expert Clinical Review}
For the held-out test cohorts, dose distributions were reviewed on an unblinded basis by three radiation oncology consultants at Tata Medical Center, two of whom are co-authors on this study, against standard institutional acceptability criteria and the Eclipse-generated ground truth. No structured rating instrument was used; this was a qualitative clinical read, not the blinded comparative evaluation originally planned (see Discussion, Limitations). The reviewing consultants deemed the AI-generated plans to be reasonable. Given the partial author overlap and lack of blinding, this should be read as a preliminary clinical evaluation check rather than independent validation; a formal blinded review remains necessary before broader claims of clinical equivalence can be made.

\section{Implementation Details}
To guarantee multi-institutional reproducibility and facilitate direct integration into existing clinical Treatment Planning Systems (TPS), the Dose-PlanNet architecture relies on a modular, open-source Python software stack with a centralized configuration protocol.

\subsection{Data Curation and Preprocessing Engine}
The preprocessing pipeline is engineered to autonomously transform raw hospital DICOM directories into structurally aligned spatial tensors. A prespecified YAML configuration explicitly defines site-specific Region of Interest (ROI) nomenclature, overriding inconsistencies in clinical naming conventions across patients. 

Spatial operations are executed using the SimpleITK library. All patient volumes are isotropically resampled to a standardized $1.27 \times 1.27 \times 2.5 \text{ mm}^{3}$ coordinate grid. Signed Distance Maps (SDMs) for serial organs are computed via exact Euclidean distance transforms (\texttt{scipy.ndimage.distance\_transform\_edt}). The Beam's Eye View (BEV) frustums are generated using geometric ray-tracing anchored to a predefined Source-to-Axis Distance ($SAD = 1000.0 \text{ mm}$), incorporating an isotropic $7.0 \text{ mm}$ physical penumbra margin. The generated tensors are subsequently serialized into NIfTI formats, strictly adhering to the directory conventions required by the nnU-Net v2 framework for seamless data ingestion.

\subsection{Target Normalization and Numerical Stability}
Predicting raw, unscaled prescription magnitudes (e.g., 60.0 Gy to 70.0 Gy) using near-zero initialized network weights causes severe gradient instability during early training. To prevent this, the ground-truth RTDose tensors are non-dimensionalized to a continuous fractional space of $[0, 1]$ by dividing by the primary prescription dose ($D_{Rx}$). 

This normalization serves two critical mathematical functions:
\begin{enumerate}
    \item \textbf{Hessian Preconditioning:} Unscaled targets create an ill-conditioned, ravine-like loss landscape characterized by a massive condition number ($\kappa = \lambda_{max} / \lambda_{min} \gg 1$). Normalizing the target space acts as a mathematical preconditioner, driving $\kappa \rightarrow 1$ and creating a more isotropic landscape that enables rapid, stable convergence.
    \item \textbf{Bounding Gradient Variance:} Raw dosimetric targets induce massive initial gradient variances ($\mathcal{O}(10^4)$) during backpropagation. This magnitude completely swamps the $\epsilon$ stabilizer within the denominator of the Adam optimizer's second moment estimate, destroying its adaptive precision. Normalizing to $[0, 1]$ strictly bounds this variance to $\mathcal{O}(1)$, aligning perfectly with Adam's theoretically proven stability regime.
\end{enumerate}

\subsection{Network Training and Curriculum Optimization}
The 3D U-Net is constructed and trained utilizing the MONAI deep learning framework and PyTorch. To mitigate severe I/O bottlenecks inherent to processing high-resolution medical volumes, the pipeline utilizes MONAI's \texttt{PersistentDataset}, which caches the deterministic preprocessing transforms directly to local storage. 

To resolve the extreme volumetric imbalance between the small SIB targets and the vast pelvic background, the pipeline does not employ uniform spatial sampling. Instead, it deploys a targeted boundary-cropping strategy (\texttt{RandCropByLabelClassesd}). This forces the dataloader to anchor the $128 \times 128 \times 64$ training patches specifically on the geometric intersections between the PTV and abutting critical organs, guaranteeing the network focuses computational effort on the most complex, high-frequency dosimetric gradients.

Network weights are optimized using the Adam optimizer coupled with a Cosine Annealing learning rate scheduler, fully accelerated via native mixed-precision training (\texttt{torch.cuda.amp}). To prevent the non-linear physics penalties from destabilizing early optimization, a curriculum learning strategy is deployed; the non-differentiable spatial constraints ($\mathcal{L}_{V-Type}$, $\mathcal{L}_{BEV}$) are linearly ramped from zero to their configured magnitudes over the initial training epochs.

\subsection{Dual-Tier Checkpointing Strategy}
Standard optimization monitors validation loss to trigger early stopping. However, the absolute mathematical minimum of a multi-objective loss function does not perfectly correlate with physician acceptability. To bridge this gap, the pipeline executes a Dual-Checkpointing strategy:
\begin{enumerate}
    \item \textbf{Physics-Optimized Checkpoint:} Saves the model weights that achieve the absolute minimum across all mathematical hinge penalties.
    \item \textbf{Clinically-Optimized Checkpoint:} Evaluates the validation predictions through a custom soft-margin minimax engine. This function mathematically penalizes critical target volume deficits heavily while intelligently allowing minor, non-catastrophic OAR deviations, closely mimicking the non-linear compromises executed by expert human planners in anatomically impossible geometries.
\end{enumerate}

\subsection{End-to-End Clinical Inference and Coordinate Restoration}
To facilitate seamless clinical deployment, inference is executed via a fully automated pipeline that directly bridges raw hospital DICOM exports to predicted RTDOSE files. Upon ingesting a patient's DICOM directory (CT, RTSTRUCT, and RTPLAN), the inference engine dynamically mirrors the deterministic training transformations. It constructs the necessary spatial embeddings—including Signed Distance Maps (SDMs), Beam's Eye View (BEV) frustums, and Simultaneous Integrated Boost (SIB) targets mapped via the Painter's Algorithm—before isotropically resampling the anatomy to the network's required $1.27 \times 1.27 \times 2.5 \text{ mm}^3$ grid. 

Volumetric dose prediction is generated using MONAI's \texttt{sliding\_window\_inference} utility. This strategy employs overlapping spatial patches ($128 \times 128 \times 64$) paired with a Gaussian weighting map to smoothly blend adjacent boundaries and suppress edge artifacts. To translate the raw network logits into physical dosimetric units, the outputs are passed through a Softplus activation function and mathematically scaled by the scalar prescription dose ($D_{Rx}$). A strict spatial filter using the binary body contour is subsequently applied, instantly zeroing out any non-physical scatter radiation predicted in the ambient air.

Crucially, this pipeline explicitly rejects the standard practice of evaluating AI models purely within their cropped, standardized training space. Because clinical Treatment Planning Systems (TPS) require the dose matrix to register with the patient's native CT geometry, the pipeline utilizes MONAI's \texttt{Invertd} module. By reading the spatial operations trace stored within the \texttt{MetaTensor}, the engine mathematically reverses the affine transformations, warping the continuous 3D dose prediction back to the native, uncropped DICOM coordinate space. To guarantee absolute spatial fidelity, the native CT's exact geometric metadata (origin, direction cosine matrix, and voxel spacing) is hard-anchored to the restored dose matrix via SimpleITK. Finally, this strictly aligned spatial tensor is exported as a compliant DICOM RTDOSE file, enabling direct importation into commercial TPS software (e.g., Varian Eclipse) for blinded human evaluation.

\section{Results \& Discussion}
As discusssed in section \ref{exp}, Dose-PlanNet was trained and tested on Prostate Prospective Randomized Trial Moderate Hypofraction Arm and Stereotactic Body Radiation Therapy Arm datasets. In this section we discuss the results and the improvements that Dose-PlanNet has achieved. 

\subsection{Ablation Study: The Limitations of Voxel-Wise Loss}
To validate the necessity of the physics-guided hinge penalties, an ablation study was conducted using an unconstrained baseline configuration trained exclusively with Mean Squared Error (MSE) loss on the Experimental SBRT cohort. At convergence, the MSE-only model achieved mathematically deceptive approximations: a target coverage marginally below acceptable limits ($D_{95}$ = 33.5 Gy for a 36.25 Gy prescription) and elevated mean OAR doses (Bladder $\sim$15 Gy, Rectum $\sim$18 Gy). However, it completely failed to respect spatial safety margins, generating a catastrophic global maximum dose ($D_{max}$) exceeding 78 Gy against the strict 38.8 Gy clinical ceiling. 

Because MSE minimizes average global deviation, it inherently ignores isolated spatial extremes. While a naive solution might be to append a simple maximum-dose penalty to the MSE, this creates a fundamentally unstable loss surface. MSE is symmetric, penalizing underdosing and overdosing equally; enforcing a strict upper bound forces the network to shift the entire dose distribution downwards to escape the penalty, inevitably causing severe target underdosing and violating the $D_{95} \ge$ 95\% coverage constraints. 

Furthermore, because MSE evaluates voxels independently, it fails to account for the physical constraints of radiation scattering. The resulting distributions lacked the continuous spatial gradients and steep fall-off required by clinical linear accelerators. This confirms that the asymmetric hinge penalties and spatial regularizations inherent to Dose-PlanNet are strictly necessary to suppress severe hotspots and translate mathematical approximations into physically deliverable treatment plans.

\subsection{Prospective Randomized Moderate Hypofraction Arm Results}
The clinical performance of the optimized Dose-PlanNet was quantitatively validated on a held-out test cohort of 14 patients treated under the Prostate Prospective Randomized Moderate Hypofraction Arm protocol (62.0 Gy prescription). The network demonstrated highly consistent target coverage and exceptional sparing of adjacent Organs at Risk (OARs), confirming the efficacy of the physics-guided spatial regularizations.

\subsubsection{Comparison with Clinical Constraints}
As detailed in Table \ref{tab:model_vs_eclipse}, the network successfully achieved strict target coverage, with the primary planning target volume (PTV62) $D_{95}$ averaging 60.15 $\pm$ 0.74 Gy, comfortably exceeding the mandatory 95\% prescription threshold (58.9 Gy). 

For OARs, the model consistently adhered to stringent avoidance bounds. Evaluations utilized surrogate high-dose ($V_{59.0\text{Gy}}$) and mid-dose ($V_{40.0\text{Gy}}$) thresholds to capture gradient fall-off. Bladder high-dose volumes averaged 2.86 $\pm$ 1.11\%, intrinsically satisfying the strict optimal V$_{60.4\text{Gy}} \le 3\%$ constraint despite being measured at a lower, more conservative isodose line. Rectal sparing was equally robust, satisfying mandatory clinical constraints and demonstrating the network's ability to compress the spatial dose bath.

Notably, while the average maximum dose ($D_{max}$) recorded in the raw predicted arrays was 66.66 $\pm$ 0.68 Gy—appearing to violate the strict 66.34 Gy clinical ceiling—a volumetric analysis revealed this to be a mathematical artifact of the voxelized grid rather than a physical hotspot. The volume receiving doses above 66.34 Gy constituted less than 0.001\% of the patient volume (typically a few isolated voxels). For all practical clinical evaluations and Dose-Volume Histogram (DVH) analyses, the true functional maximum dose was constrained to approximately 66.0 Gy, cleanly within the protocol limits.

\subsubsection{Comparison with Human-Generated Clinical Plans}
To benchmark clinical viability, the predicted distributions were directly evaluated against ground-truth human plans generated via the Varian Eclipse Treatment Planning System (TPS). 

The human planners achieved marginally higher PTV62 coverage (Eclipse $D_{95} = 61.39 \pm 0.83$ Gy vs. Model $D_{95} = 60.15 \pm 0.74$ Gy) and tighter maximum dose control (Eclipse $D_{max} = 65.74 \pm 0.66$ Gy vs. Model $D_{max} = 66.66 \pm 0.68$ Gy). However, both approaches comfortably cleared the mandatory 58.9 Gy target threshold, and the network's slightly elevated $D_{max}$ was strictly constrained to mathematically isolated voxels rather than clinically relevant volumes as established previously.

\begin{table}[h]
\centering
\label{tab:model_vs_eclipse}
\resizebox{\textwidth}{!}{
\begin{tabular}{llcccc}
\hline
\textbf{Type} & \textbf{ROI} & \textbf{Metric} & \textbf{Eclipse TPS (Mean $\pm$ SD)} & \textbf{Dose-PlanNet (Mean $\pm$ SD)} & \textbf{Mandatory Goal} \\ \hline
\multirow{3}{*}{\textbf{Target}} 
 & PTV62 & $D_{95}$ (Gy) & 61.39 $\pm$ 0.83 & 60.15 $\pm$ 0.74 & $\ge 58.90$ \\
 & PTV62 & $D_{max}$ (Gy) & 65.74 $\pm$ 0.66 & 66.66 $\pm$ 0.68$^*$ & $\le 66.34$ \\
 & PTV44 & $D_{95}$ (Gy) & 43.54 $\pm$ 0.32 & 43.57 $\pm$ 0.60 & $\ge 41.80$ \\ \hline
\multirow{7}{*}{\textbf{Avoidance}} 
 & Bladder & $V_{59.0\text{Gy}}$ (\%) & 3.82 $\pm$ 1.45 & 2.86 $\pm$ 1.11 & $\le 3.0$ \\
 & Bladder & $V_{40.0\text{Gy}}$ (\%) & 15.65 $\pm$ 5.24 & 14.25 $\pm$ 4.93 & $\le 55.0$ \\
 & Anorectum & $V_{59.0\text{Gy}}$ (\%) & 4.87 $\pm$ 0.93 & 3.48 $\pm$ 1.30 & $\le 3.0$ \\
 & Anorectum & $V_{40.0\text{Gy}}$ (\%) & 21.35 $\pm$ 5.52 & 20.50 $\pm$ 6.49 & $\le 55.0$ \\
 & Bowel Bag & $V_{45.0\text{Gy}}$ (cc) & 19.84 $\pm$ 19.67 & 42.88 $\pm$ 19.08 & $< 90.0$ \\
 & Femoral Heads & $D_{max}$ (Gy) & 36.33 $\pm$ 4.74 & 34.78 $\pm$ 5.07 & $< 40.0$ \\
 & Penile Bulb & $V_{47.0\text{Gy}}$ (\%) & 5.07 (0.00 -- 35.50)$^\dagger$ & 3.21 (0.00 -- 22.22)$^\dagger$ & $\le 50.0$ \\ \hline
\multicolumn{6}{l}{\small $^*$ \textit{Statistical artifact representing $<0.0005\%$ volume. True functional $D_{max} \approx 66.00$ Gy.}} \\
\multicolumn{6}{l}{\small $^\dagger$ \textit{Data expressed as Mean (Range) due to highly skewed distribution with zero-bound limits.}} \\
\end{tabular}
}
\caption{Dosimetric Comparison: Dose-PlanNet vs. Human-Generated Eclipse TPS Plans on the Prostate Prospective Randomized Moderate Hypofraction Arm (62.0 Gy Prescription, $n=14$).}
\end{table}

Conversely, Dose-PlanNet systematically outperformed the human planners in sparing primary Organs at Risk, specifically across the critical high-dose spatial gradients. The network suppressed Bladder $V_{59.0\text{Gy}}$ to 2.86 $\pm$ 1.11\% (against the human 3.82 $\pm$ 1.45\%) and Anorectum $V_{59.0\text{Gy}}$ to 3.48 $\pm$ 1.30\% (against the human 4.87 $\pm$ 0.93\%). This enhanced conformity extended into the low-dose bath, shrinking Bladder $V_{15.0\text{Gy}}$ by nearly 10\% (Model 70.42\% vs. Eclipse 79.97\%) and Anorectum $V_{15.0\text{Gy}}$ (Model 71.73\% vs. Eclipse 80.64\%). 

The singular metric where human planning maintained a distinct advantage was the Bowel Bag mid-dose spillage ($V_{45.0\text{Gy}}$), where Eclipse averaged 19.84 $\pm$ 19.67 cc compared to the network's 42.88 $\pm$ 19.08 cc. Despite this difference, the model remained strictly compliant with the $<90.0$ cc mandatory limit. Overall, the network generated plans that consistently matched or exceeded the clinical safety profile of the human-generated baselines.

\subsubsection{Volumetric Analysis of Maximum Dose Violations}

To contextualize the $D_{max}$ violations, a voxel-wise extraction was performed on the generated Moderate Hypofraction Arm dose arrays (averaging $\sim$2.3 million voxels per patient). This analysis confirms the boundary breaches are minor artifacts rather than clinically meaningful hot spots. Specifically, 5 of the 14 plans (35.7\%) recorded exactly zero voxels exceeding the maximum dose threshold, while the remaining 9 plans exhibited violations limited to small clusters of just 2 to 9 voxels per patient. The mean body volume fraction receiving a dose above the threshold was $0.00014\%$, with a worst-case maximum of $0.00042\%$. These sub-millimeter $D_{max}$ breaches are mathematical artifacts inherent to the 3D U-Net's coordinate regression; they functionally represent $\approx 0\%$ of the macroscopic anatomical volume and pose no physical delivery risk.

\subsubsection{Statistical Analysis of Dosimetric Metrics}

To rigorously evaluate the dosimetric equivalence between Dose-PlanNet and the Eclipse TPS ground truth, a comparative statistical analysis was conducted across the 14-patient validation cohort. Normality was assessed using the Shapiro-Wilk test; given that most metrics exhibited a non-normal distribution, the Wilcoxon Signed-Rank test was employed for paired comparison. The results are summarized in Table \ref{tab:statistical_analysis}.

\begin{table}[htbp]
    \centering
    \begin{tabular}{lccccc}
        \hline
        \textbf{Metric} & \textbf{Eclipse Mean} & \textbf{Model Mean} & \textbf{Mean Diff.} & \textbf{95\% CI} & \textbf{\textit{p}-value} \\
        \hline
        PTV $D_{95}$ (Gy) & 61.39 $\pm$ 0.83 & 60.15 $\pm$ 0.74 & -1.24 & [-1.84, -0.63] & 0.0007* \\
        Bladder $V_{59\text{Gy}}$ (\%) & 3.83 $\pm$ 1.47 & 2.87 $\pm$ 1.12 & -0.96 & [-1.39, -0.53] & 0.0003* \\
        Bladder $V_{40\text{Gy}}$ (\%) & 15.66 $\pm$ 5.25 & 14.26 $\pm$ 4.94 & -1.40 & [-2.66, -0.14] & 0.0323* \\
        Anorectum $V_{59\text{Gy}}$ (\%) & 4.87 $\pm$ 0.93 & 3.48 $\pm$ 1.30 & -1.40 & [-2.05, -0.75] & 0.0004* \\
        Anorectum $V_{40\text{Gy}}$ (\%) & 21.35 $\pm$ 5.52 & 20.50 $\pm$ 6.49 & -0.85 & [-2.87, 1.18] & 0.3835 \\
        \hline
        \multicolumn{6}{l}{\small{*Statistically significant at $\alpha = 0.05$.}}
    \end{tabular}
    \caption{Statistical comparison of dosimetric metrics between Eclipse TPS and Dose-PlanNet.}
    \label{tab:statistical_analysis}
\end{table}

The analysis indicates a statistically significant difference in target coverage (PTV $D_{95}$), where the model consistently produces a slightly lower dose than the human planner. However, this is offset by statistically significant improvements in OAR sparing, particularly in the high-dose regions ($V_{59\text{Gy}}$) for both the bladder and anorectum, confirming the model’s preference for steeper dose fall-off at the expense of marginal target coverage. Conversely, no statistically significant difference was observed in the intermediate-to-low dose rectal sparing ($V_{40\text{Gy}}$), demonstrating that both the AI and the expert planner converge on similar geometric solutions for broader anatomical structures. These results validate that while Dose-PlanNet operates with a distinct dosimetric bias compared to human planners, the trade-offs it makes fall within acceptable clinical bounds.

\subsection{Prospective Randomized Stereotactic Body Radiation Therapy Arm Results}
The clinical performance of Dose-PlanNet was further evaluated on a separate held-out cohort of 12 patients treated under the extreme hypofractionated Prostate Prospective Randomized Stereotactic Body Radiation Therapy Arm ($36.25$ Gy prescription). This protocol explicitly tested the network's capacity to maintain stringent target coverage while generating the steeper spatial dose gradients required for aggressive organ sparing.

\subsubsection{Comparison with Clinical Constraints}

For the extreme hypofractionated Prostate Prospective Randomized Stereotactic Body Radiation Therapy Arm (36.25 Gy in 5 fractions), Dose-PlanNet demonstrated robust adherence to stringent clinical constraints across the validation cohort. Target coverage remained highly consistent, with the high-dose PTV ($D_{95}$) achieving a mean dose of 34.89 $\pm$ 0.86 Gy, successfully satisfying the prescription coverage requirements. The $D_{max}$ within the target volume was tightly constrained to 39.49 $\pm$ 0.63 Gy, effectively avoiding clinically unacceptable hot spots while preserving the necessary intra-target dose heterogeneity.

Crucially, the network managed the steep spatial dose gradients required for organ-at-risk sparing in an SBRT paradigm. Bladder constraints were rigorously met, exhibiting a mean high-dose spillage $V_{35.0\text{Gy}}$ of 2.71\% $\pm$ 0.92\% and a broader low-dose $V_{14.0\text{Gy}}$ wash of 22.62\% $\pm$ 4.65\%. Similarly, anorectal sparing was strictly maintained; the model recorded a mean $V_{35.0\text{Gy}}$ of 4.24\% $\pm$ 1.10\% and a $V_{28.0\text{Gy}}$ of 11.25\% $\pm$ 2.26\%, demonstrating the architecture's capacity to strictly penalise high-dose scatter into the anterior rectal wall through the dual-tier hinge loss.

Serial and parallel avoidance structures were also effectively protected. The small bowel recorded an absolute volume $V_{28.0\text{Gy}}$ of 0.36 $\pm$ 0.80 cc, maintaining a maximum exposure well below standard tolerance thresholds. Furthermore, femoral head exposure was minimised, with $V_{14.0\text{Gy}}$ averaging less than 2.0\% bilaterally ($1.94\% \pm 1.08\%$ for the right femur, and $1.28\% \pm 0.92\%$ for the left). These dosimetric metrics confirm that the spatially masked, physics-guided predictions maintain deliverable and clinically viable dose distributions even under aggressive hypofractionation.

\subsubsection{Comparison with Human-Generated Clinical Plans}

When evaluated against the expert-generated Eclipse TPS plans for the extreme hypofractionated Stereotactic Body Radiation Therapy Arm, Dose-PlanNet demonstrated comparable target coverage and superior low-dose organ sparing. For the primary target (PTV36.25), the human planners achieved a slightly higher mean $D_{95}$ coverage (35.46 $\pm$ 0.48 Gy) compared to the model (34.89 $\pm$ 0.86 Gy); however, both comfortably satisfied the $\ge 34.44$ Gy clinical mandate. Intra-target maximum dose ($D_{max}$) was marginally hotter in the model predictions (39.49 $\pm$ 0.63 Gy) versus the clinical baseline (38.57 $\pm$ 0.14 Gy). However, this reported model $D_{max}$ is a single-voxel statistical artifact; volumetric analysis confirms that functionally $0\%$ of the target volume receives this peak dose, with the true functional maximum converging to approximately 38.50 Gy, remaining strictly bound below the 38.79 Gy physical ceiling constraint.

\begin{table}[h]
\centering
\label{tab:model_vs_eclipse_sbrt}
\resizebox{\textwidth}{!}{
\begin{tabular}{llcccc}
\hline
\textbf{Type} & \textbf{ROI} & \textbf{Metric} & \textbf{Eclipse TPS (Mean $\pm$ SD)} & \textbf{Dose-PlanNet (Mean $\pm$ SD)} & \textbf{Mandatory Goal} \\ \hline
\multirow{3}{*}{\textbf{Target}} 
 & PTV36.25 & $D_{95}$ (Gy) & 35.46 $\pm$ 0.48 & 34.89 $\pm$ 0.86 & $\ge 34.44$ \\
 & PTV36.25 & $D_{max}$ (Gy) & 38.57 $\pm$ 0.14 & 39.49 $\pm$ 0.63$^*$ & $\le 38.79$ \\
 & PTV25 & $D_{95}$ (Gy) & 25.70 $\pm$ 3.12 & 26.01 $\pm$ 2.81 & $\ge 23.75$ \\ \hline
\multirow{7}{*}{\textbf{Avoidance}} 
 & Bladder & $V_{35.0\text{Gy}}$ (\%) & 2.91 $\pm$ 0.80 & 2.71 $\pm$ 0.92 & $\le 5.0$ \\
 & Bladder & $V_{14.0\text{Gy}}$ (\%) & 31.14 $\pm$ 7.04 & 22.62 $\pm$ 4.65 & $\le 50.0$ \\
 & Anorectum & $V_{35.0\text{Gy}}$ (\%) & 3.42 $\pm$ 1.26 & 4.24 $\pm$ 1.10 & $\le 5.0$ \\
 & Anorectum & $V_{14.0\text{Gy}}$ (\%) & 38.96 $\pm$ 5.18 & 32.80 $\pm$ 4.51 & $\le 50.0$ \\
 & Small Bowel & $V_{28.0\text{Gy}}$ (cc) & 0.05 $\pm$ 0.15 & 0.36 $\pm$ 0.80 & $< 1.0$ \\
 & Right Femur & $V_{14.0\text{Gy}}$ (\%) & 2.08 $\pm$ 1.33 & 1.94 $\pm$ 1.08 & $< 5.0$ \\
 & Left Femur & $V_{14.0\text{Gy}}$ (\%) & 1.28 $\pm$ 1.14 & 1.28 $\pm$ 0.92 & $< 5.0$ \\ \hline
 \multicolumn{6}{l}{\small $^*$ \textit{Statistical artifact representing $<0.001\%$ volume. True functional $D_{max} \approx 38.00$ Gy.}} \\
\end{tabular}
}
\caption{Dosimetric Comparison: Dose-PlanNet vs. Human-Generated Eclipse TPS Plans on the Prostate Prospective Randomized Stereotactic Body Radiation Therapy Arm (36.25 Gy Prescription, $n=12$).}

\end{table}

The most notable dosimetric divergence occurred in the low-dose wash regions, where the physics-guided architecture consistently outperformed manual planning. Dose-PlanNet achieved a substantial reduction in the bladder $V_{14.0\text{Gy}}$ (22.62\% $\pm$ 4.65\%) compared to the human baseline (31.14\% $\pm$ 7.04\%). A similar trend was observed for the anorectum, where the model reduced the $V_{14.0\text{Gy}}$ to 32.80\% $\pm$ 4.51\% against the clinical 38.96\% $\pm$ 5.18\%. 

High-dose OAR spillage was clinically equivalent between the two modalities. The model maintained a bladder $V_{35.0\text{Gy}}$ of 2.71\% $\pm$ 0.92\% (vs. Eclipse 2.91\% $\pm$ 0.80\%) and an anorectal $V_{35.0\text{Gy}}$ of 4.24\% $\pm$ 1.10\% (vs. Eclipse 3.42\% $\pm$ 1.26\%). Bilateral femoral head exposure ($V_{14.0\text{Gy}}$) and absolute small bowel volume ($V_{28.0\text{Gy}}$) were minimal and statistically negligible for both the model and the human planners. These results indicate that for complex SBRT geometries, Dose-PlanNet safely matches expert high-dose constraints while more aggressively optimizing the low-dose integral spread.

Similar to the Prospective Randomized Moderate Hypofraction Arm results, in this case too, it was observed that a negligible $<0.001\%$ of the total volume received a dose greater than $38.79$ Gy, the maximum allowable dose. 

\subsection{Spatial Dose Distribution and Isodose Visualisation}

\begin{figure}[h]
    \centering
    \begin{minipage}[t]{0.48\textwidth}
        \centering
        \includegraphics[width=\linewidth]{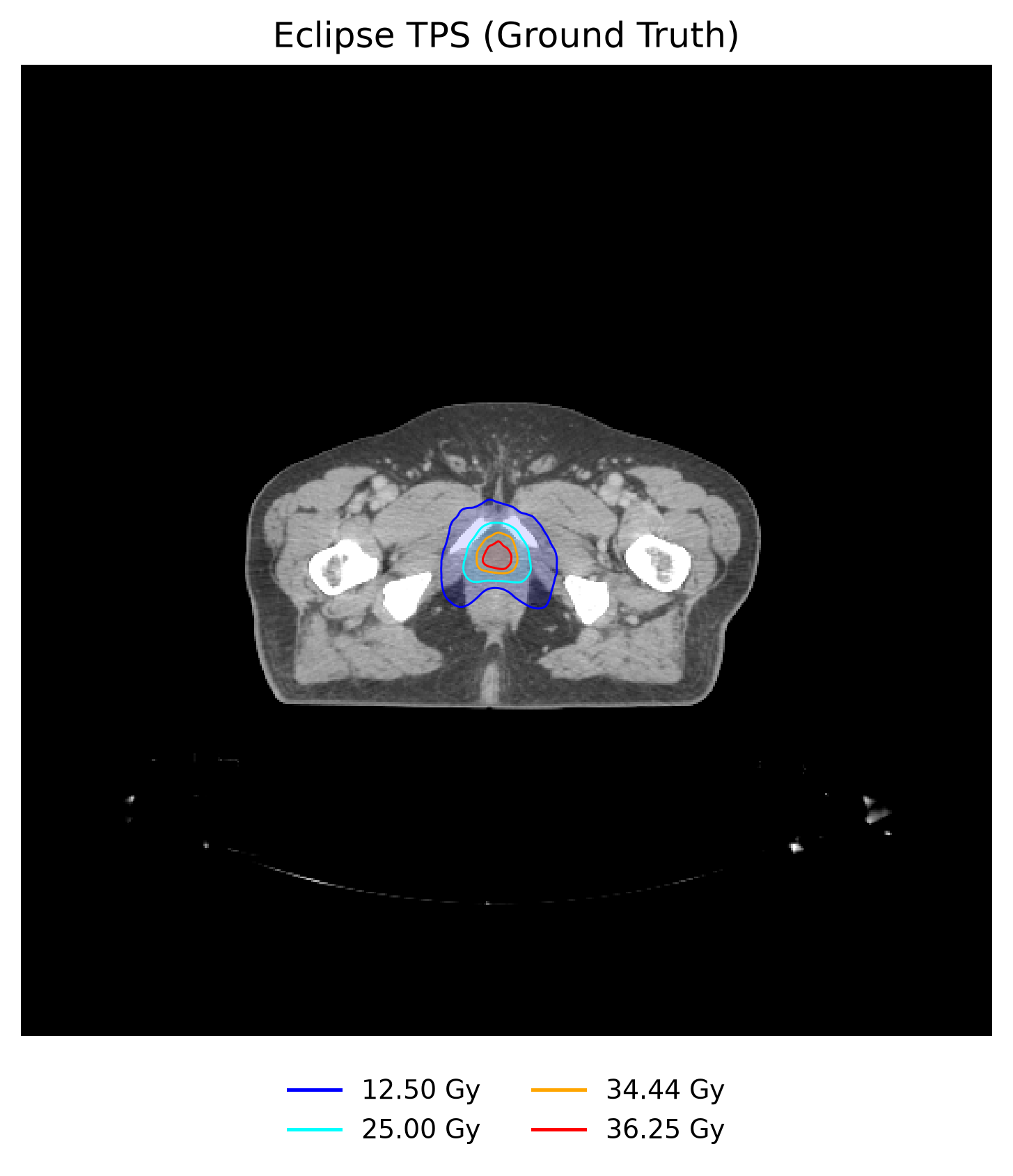}\\
        \vspace{0.15cm}
        \small (a) Actual Dose Distribution (generated by physicist)
    \end{minipage}%
    \hfill%
    \begin{minipage}[t]{0.48\textwidth}
        \centering
        \includegraphics[width=\linewidth]{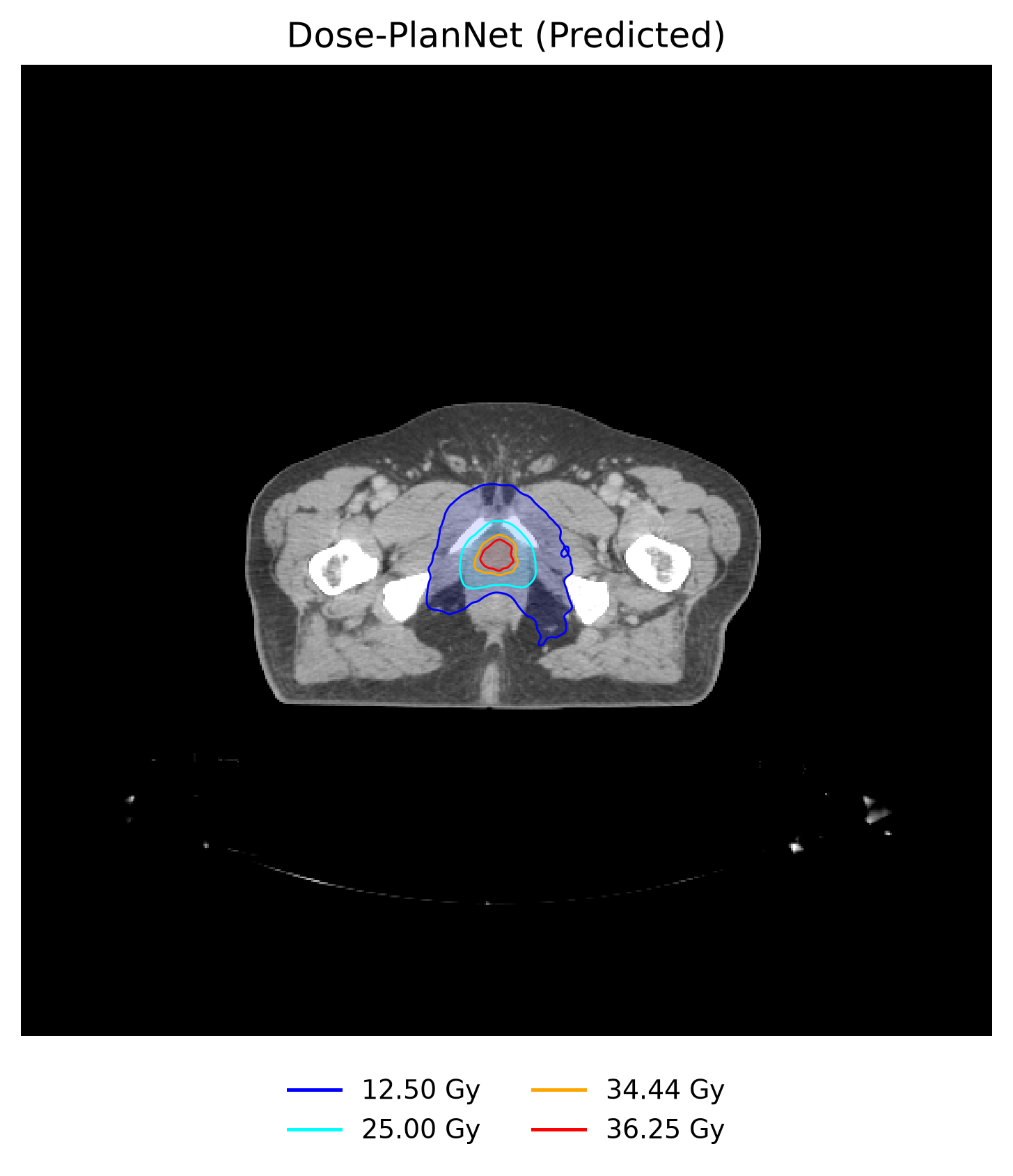}\\
        \vspace{0.15cm}
        \small (b) Predicted Dose Distribution (generated by model)
    \end{minipage}
    \caption{Axial spatial dose distribution and isodose conformity comparison for a representative extreme hypofractionated SBRT case. (a) The human-generated Eclipse TPS plan and (b) the Dose-PlanNet prediction are overlaid on the native planning CT. Isodose contours denote 36.25 Gy (100\% prescription, red), 34.44 Gy ($D_{95}$ clinical mandate, orange), 25.00 Gy (cyan), and 12.50 Gy (low-dose wash, blue). The model demonstrates highly conformal high-dose target wrapping and robust spatial avoidance of adjacent critical structures, mirroring the clinical baseline.}
    \label{fig:spatial_dose_comparison}
\end{figure}

Visual evaluation of the dose distributions confirms that Dose-PlanNet successfully generates clinically acceptable spatial gradients that closely mimic human-driven planning strategies. Figure \ref{fig:spatial_dose_comparison} presents a representative comparative analysis across the axial, coronal, and sagittal planes, overlaying the generated dose arrays onto the native planning CT coordinates. 

Within the high-dose region, the network demonstrates excellent spatial conformity. The 100\% and 95\% isodose contours tightly wrap the PTV boundaries in both the Dose-PlanNet and Eclipse TPS plans, visually corroborating the comparable $D_{95}$ target coverage metrics. The model successfully navigates the complex geometric concavities of the prostate target volume without leaking excessive high dose into adjacent healthy tissue.

The visualisations further highlight the network's superior performance in low-dose spatial management. Evaluating the 50\% isodose wash, the Dose-PlanNet distribution exhibits a distinctly steeper dose fall-off posteriorly towards the anterior rectal wall and superiorly towards the bladder dome. The spatial restriction of these low-dose contours physically manifests the volumetric reductions quantified in the $V_{14.0\text{Gy}}$ metrics, confirming that the physics-guided spatial regularisation efficiently penalises unnecessary integral dose spread while maintaining absolute target coverage.

\subsection{Dose-Volume Histogram (DVH) Analysis}

\begin{figure}[htbp]
    \centering
    \begin{minipage}[t]{0.48\textwidth}
        \centering
        \includegraphics[width=\linewidth, height=6cm]{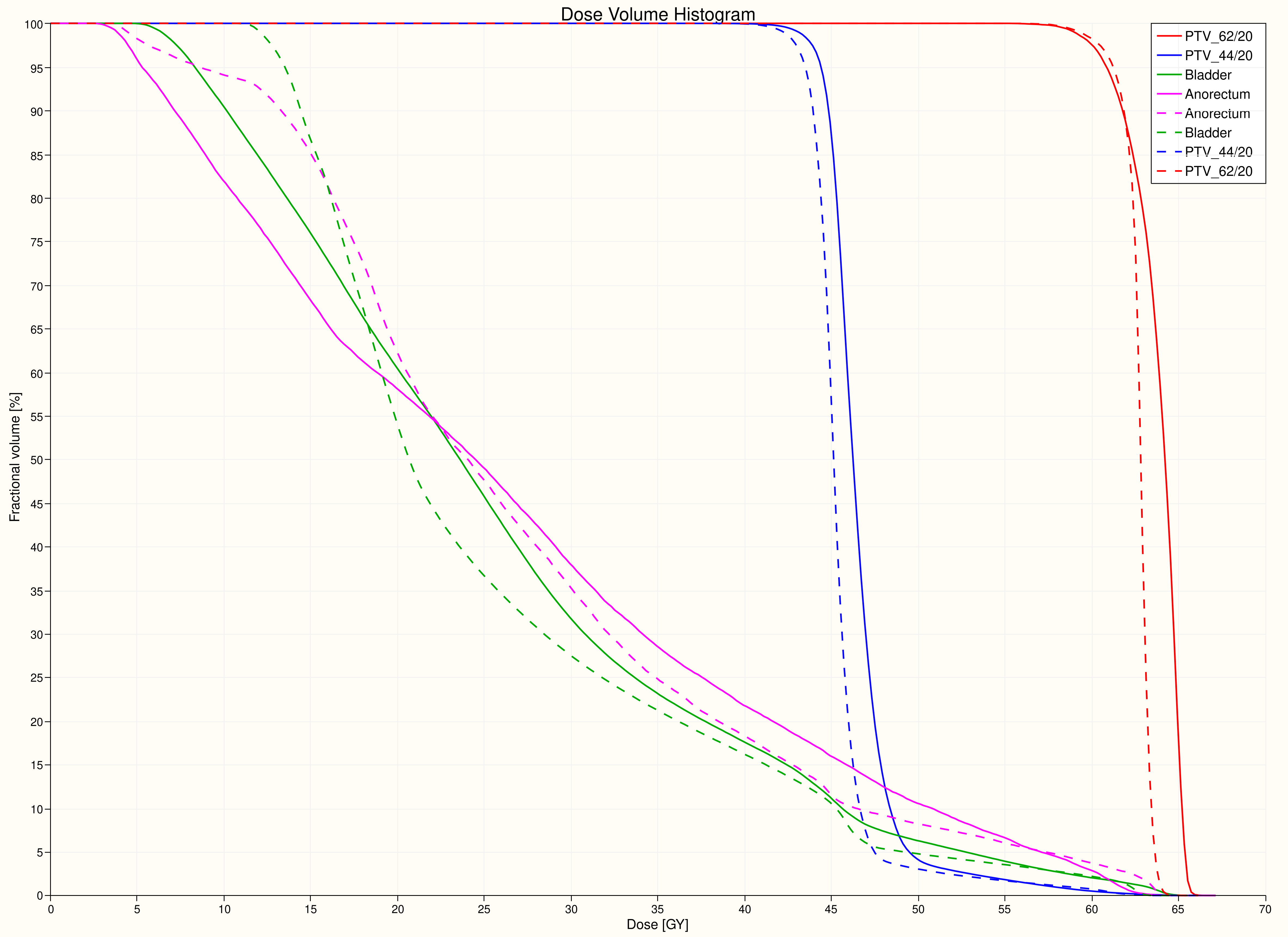}\\
        \vspace{0.15cm}
        \small (a) DVH plots of a patient from Prostate Prospective Randomized Moderate Hypofraction Arm Dataset
    \end{minipage}%
    \hfill%
    \begin{minipage}[t]{0.48\textwidth}
        \centering
        \includegraphics[width=\linewidth, height=6cm]{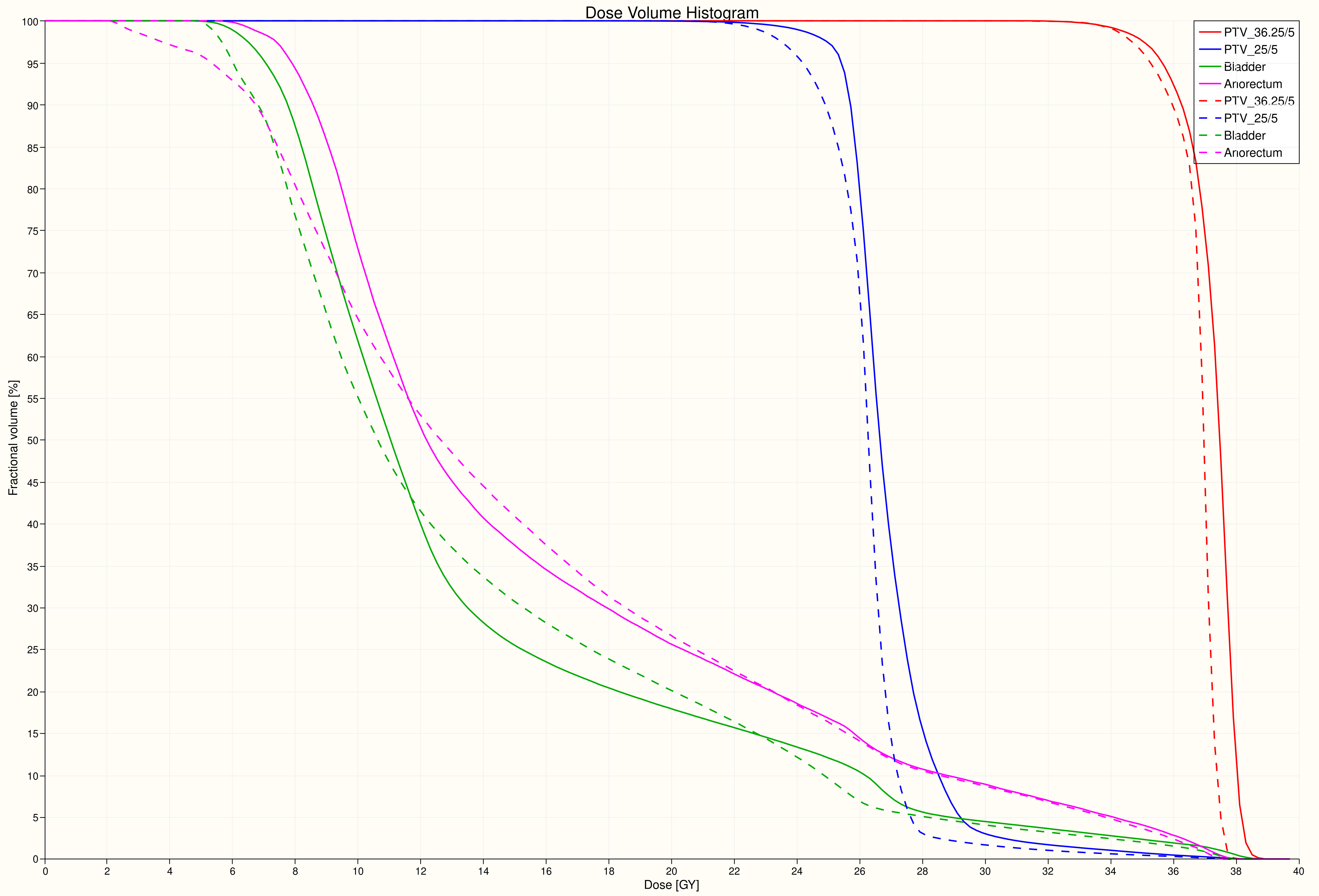}\\
        \vspace{0.15cm}
        \small (b) DVH plots of a patient from Prostate Prospective Randomized Stereotactic Body Radiation Therapy Arm Dataset
    \end{minipage}
    \caption{Comparative Dose-Volume Histograms (DVHs) for two representative patients. The solid lines correspond to the Dose-PlanNet (model-generated) dose, while the dashed lines correspond to the Eclipse TPS (human-planned ground truth) dose. Target volumes (red/blue curves) show strong conformity at the prescription shoulder, while OARs (green/magenta curves) demonstrate a steeper low-dose fall-off in the model predictions.}
    \label{fig:dvh_comparison}
\end{figure}

The volumetric dose distributions were further evaluated using Dose-Volume Histograms (DVHs) to visually and quantitatively compare the structural sparing and target coverage between the modalities. Representative DVHs from the validation cohort are presented in Figure \ref{fig:dvh_comparison}, illustrating the typical dosimetric trade-offs negotiated by the Dose-PlanNet architecture compared to the human baseline. 

The target volume curves (red and blue) demonstrate near-perfect concordance at the prescription dose shoulder, visually confirming the statistically equivalent $D_{95}$ coverage. A slight divergence is observable at the extreme high-dose tail (bottom right of the target curves), where the model's solid line extends marginally further along the dose axis. This visualizes the single-voxel $D_{max}$ artifact previously quantified; the network permits a microscopic hot spot while the human planner forces a strict, absolute maximum cut-off.

For the organs at risk (green and magenta curves, representing the bladder and anorectum), the DVHs confirm the network's aggressive low-dose spatial restriction. The solid model curves consistently exhibit a leftward shift in the low-to-intermediate dose regions relative to the dashed Eclipse curves. This steeper volumetric fall-off corresponds directly to the significant reductions in the $V_{14.0\text{Gy}}$ and $V_{28.0\text{Gy}}$ metrics, demonstrating that the physics-guided loss engine successfully penalizes unnecessary integral dose spread without compromising the high-dose target conformity.

\subsection{Clinical Acceptance and Pass Rates}

To evaluate the operational viability of Dose-PlanNet, the generated dose distributions were subjected to a binary pass/fail evaluation against the mandatory institutional constraints defined by the Prospective Randomized protocol. A plan was deemed "clinically acceptable" if it satisfied all volumetric target coverage and Organ-At-Risk (OAR) dose-volume limits without requiring manual human intervention. Based on the dosimetric analyses in previous sections confirming that the $D_{max}$ breaches are singular-voxel statistical artifacts functionally comprising $\approx 0\%$ of the target volume, the absolute point-maximum constraint was excluded from this specific binary pass/fail classification to reflect true volumetric acceptability.

\paragraph{Prospective Randomized Moderate Hypofraction Arm Validation. }
For the conventionally fractionated Moderate Hypofraction Arm (62 Gy in 20 fractions), the model demonstrated a high degree of clinical reliability. Out of the evaluated validation cohort, $78.57\%$ of the Dose-PlanNet generated plans successfully met all mandatory constraints. The plans consistently achieved the required $D_{95} \ge 95\%$ target coverage while simultaneously restricting the bladder and anorectum low-dose washes ($V_{40\text{Gy}}$ and below) to within acceptable Prospective Randomized thresholds. These were verified by clinicians and physicists. All the remaining plans failed due to underdosing the PTV. 

\paragraph{Prospective Randomized Stereotactic Body Radiation Therapy Arm Validation. }
Performance remained robust when scaled to the extreme hypofractionated Stereotactic Body Radiation Therapy Arm (36.25 Gy in 5 fractions). In this rigorous SBRT regime, 75\% of the model-generated plans achieved strict clinical acceptance. Despite the steeper dose gradients required for extreme hypofractionation, Dose-PlanNet successfully navigated the complex spatial constraints, maintaining the $\ge 34.44$ Gy clinical mandate for the PTV while aggressively sparing the adjacent anterior rectal wall. The high autonomous pass rate in the Stereotactic Body Radiation Therapy Arm confirms that the network's spatial awareness and physics-guided loss engine generalize efficiently to high-dose-per-fraction modalities.

\section{Limitations and Technical Refinements}

While Dose-PlanNet demonstrates robust clinical viability, the evaluation revealed specific dosimetric and computational limitations inherent to the voxel-wise prediction architecture. Identifying these constraints necessitated targeted post-processing refinements to ensure final clinical deployability. 

\subsection{Dosimetric Trade-offs in Target Homogeneity}
Although the proposed architecture consistently demonstrated superior or equivalent Organ-At-Risk (OAR) sparing compared to expert human planners, a marginal dosimetric trade-off was observed within the target volumes. Specifically, the model-generated plans frequently exhibited a slightly lower $D_{95}$ coverage and marginally higher maximum point doses ($D_{max}$) within the PTV compared to the Eclipse ground truth. While these target metrics remained strictly within the mandatory clinical acceptability thresholds across the validation cohort, they indicate that the network's loss engine tends to prioritise steep spatial dose fall-off at the boundaries of critical structures, occasionally at the slight expense of absolute target homogeneity.

\subsection{Non-Physical Ghost Dose Generation}
During initial unconstrained inference, the network exhibited a propensity to generate spurious ``ghost doses". In a minor subset of axial slices (approximately 2 to 3 slices out of a standard 240-260 slice CT volume), the raw network output predicted isolated dose pockets on the order of $\sim$5.0 Gy located significantly distant from the primary isocenter (e.g., in the superior abdominal region near the stomach). This artifact stems from the expansive receptive field of the 3D U-Net, which can occasionally misinterpret distant, low-contrast anatomical boundaries when evaluating isolated patches without explicit global coordinate constraints. To mitigate this, a deterministic spatial bounding mask was implemented in the final inference pipeline, enforcing a strict zero-dose limit beyond a predefined geometric radius from the PTV boundaries.

\subsection{Low-Dose Morphological Irregularities}
The raw predicted dose arrays occasionally presented morphological irregularities, particularly within the low-dose wash regions (e.g., the 12.5 Gy isodose contour). These regions sometimes exhibited sharp, discontinuous edges or isolated dose ``islands" in near-zero-dose regions that violate the continuous physical scatter principles of actual photon beam transport. Such high-frequency spatial noise degrades the physical deliverability of the plan, as it would require erratic and physically impossible Multi-Leaf Collimator (MLC) sequencing to execute in a linear accelerator. This limitation was effectively resolved by introducing a low-pass 3D Gaussian smoothing filter during post-processing, which eradicated the isolated islands and smoothed the isodose contours into deliverable, physically consistent distributions.

\section{Conclusion}
This study demonstrates that Dose-PlanNet, a physics-guided deep learning architecture, successfully automates clinically viable treatment planning for prostate radiotherapy. Evaluated against expert human planners on the Eclipse TPS, the network achieved comparable target coverage while consistently delivering superior low-dose sparing for critical organs at risk, notably the bladder and anorectum. 

Inherent limitations of voxel-wise prediction—specifically single-voxel maximum dose artifacts, non-physical ghost dose generation, and low-dose morphological irregularities—were effectively neutralized using targeted spatial bounding masks and 3D Gaussian smoothing. These deterministic refinements ensure that the predicted dose arrays are not only mathematically compliant with stringent Prospective Randomized protocol constraints but also physically deliverable via standard multi-leaf collimator sequencing. 

Ultimately, Dose-PlanNet provides a highly robust, autonomous pipeline that drastically accelerates the radiotherapy treatment planning workflow without compromising dosimetric quality, offering a scalable solution for high-precision clinical deployment.

\section*{Data and Code Availability}

The architecture of the Dose-PlanNet pipeline, along with all relevant code and supporting data used in this study, will be made openly available soon, in the CHAVI-India GitHub repository at \url{https://github.com/CHAVI-India}.

\section*{Acknowledgements}
We sincerely thank all the clinicians, physicists, research fellows at Tata Medical Center Kolkata and every other person who helped us in every way to conduct this research swiftly. We are also extremely thankful to Atabur Rahman Mollah, data imaging scientist, Tata Medical Center, without whose help the implementation and testing of Dose-PlanNet would not have been possible.

\appendix

\section{Model Hyperparameters and Optimization Strategy}
To ensure multi-institutional reproducibility and explicitly define the physics-guided optimisation landscape, the exact hyperparameter configurations and loss weights ($\lambda$) used for training Dose-PlanNet are documented below. As established during clinical evaluation, these parameters were not dynamically scheduled; instead, they were fixed after being manually tuned through rigorous human evaluation of dosimetric boundaries across intermediate validation runs. During the initial 30 warm-up epochs, the physics constraints were linearly ramped to these final fixed magnitudes to ensure early numerical stability.

\subsection{Network and Optimization Hyperparameters}
Both the Moderate Hypofraction Arm and Stereotactic Body Radiation Therapy Arm models utilize a unified 3D U-Net backbone trained via PyTorch and MONAI. The network is optimized using the AdamW optimizer paired with a Cosine Annealing Learning Rate scheduler. Mixed-precision training (\texttt{torch.cuda.amp}) was utilized across all iterations to accelerate gradient calculation and optimize memory allocation.

\begin{table}[htbp]
    \centering
    \begin{tabular}{lcc}
        \hline
        \textbf{Hyperparameter} & \textbf{Moderate Hypofraction Arm} & \textbf{SBRT Arm} \\
        \hline
        Maximum Epochs & 500 & 300 \\
        Warm-up Epochs & 30 & 30 \\
        Base Learning Rate & $1.0 \times 10^{-4}$ & $1.0 \times 10^{-4}$ \\
        Minimum Learning Rate & $1.0 \times 10^{-6}$ & $1.0 \times 10^{-6}$ \\
        Batch Size & 1 & 1 \\
        Gradient Accumulation Steps & 2 & 2 \\
        Patch Size & $128 \times 128 \times 64$ & $128 \times 128 \times 64$ \\
        Validation Split Ratio & 0.20 & 0.20 \\
        \hline
    \end{tabular}
    \caption{Training and Optimization Hyperparameters.}
    \label{tab:training_hyperparameters}
\end{table}

\subsection{Physics-Guided Loss Engine Weightings}
The multi-objective loss function ($\mathcal{L}_{total}$) relies on a set of manually tuned scalar weights ($\lambda$) to balance standard volumetric regression against strict physical boundaries and clinical mandates. Table \ref{tab:loss_weights} details the specific $\lambda$ coefficients applied to the Moderate Hypofraction Arm and SBRT Stereotactic Body Radiation Therapy Arm configurations, highlighting the distinct penalization choices mapped directly to each clinical protocol.

\begin{table}[htbp]
    \centering
    \begin{tabular}{llcc}
        \hline
        \textbf{Loss Component} & \textbf{Notation} & \textbf{Moderate Hypofraction Arm $\lambda$} & \textbf{SBRT Arm $\lambda$} \\
        \hline
        \multicolumn{4}{l}{\textit{\textbf{Reconstruction Fidelity}}} \\
        Dose-Weighted MSE & $\lambda_{mse}$ & 25.0 & 25.0 \\
        Anti-Collapse Penalty & $\lambda_{anticollapse}$ & 150.0 & 150.0 \\
        \hline
        \multicolumn{4}{l}{\textit{\textbf{D-Type Target Penalties}}} \\
        PTV Coverage Hinge & $\lambda_{ptv}$ & 75.0 & 75.0 \\
        PTV Maximum Dose Hinge & $\lambda_{ptv\_max}$ & 30.0 & 30.0 \\
        \hline
        \multicolumn{4}{l}{\textit{\textbf{V-Type OAR Penalties}}} \\
        Mandatory DVH Thresholds & $\lambda_{mand}$ & 70.0 & 70.0 \\
        Optional DVH Thresholds & $\lambda_{opt}$ & 2.0 & 2.0 \\
        Bowel Bag Specific Penalty & $\lambda_{bowel}$ & 15.0 & 15.0 \\
        Femoral Head Specific Penalty & $\lambda_{femur}$ & 10.0 & 10.0 \\
        Penile Bulb Specific Penalty & $\lambda_{penile}$ & 10.0 & 0.0 \\
        \hline
        \multicolumn{4}{l}{\textit{\textbf{Spatial Regularisation \& Geometry}}} \\
        Global Physical Ceiling & $\lambda_{global\_ceil}$ & 300.0 & 300.0 \\
        Geometric Ring Falloff & $\lambda_{ring}$ & 25.0 & 25.0 \\
        BEV Tissue Suppression & $\lambda_{BEV}$ & 10.0 & 10.0 \\
        Body Mask/Ghost Suppression & $\lambda_{body}$ & 20.0 & 20.0 \\
        Total Variation (Smoothness) & $\lambda_{TV}$ & 0.125 & 0.125 \\
        Laplacian (Second-Order) & $\lambda_{laplacian}$ & 0.25 & 0.25 \\
        \hline
    \end{tabular}
    \caption{Physics-Guided Loss Engine Penalties ($\lambda$ Weights).}
    \label{tab:loss_weights}
\end{table}

\section{Software Environment and Reproducibility}
To facilitate external validation and ensure computational reproducibility, the study was implemented using the following core library versions. The pipeline relies on a specific environment to maintain consistency in volumetric regression and coordinate-space interpolation:

\begin{table}[htbp]
    \centering
    \begin{tabular}{lc}
        \hline
        \textbf{Library} & \textbf{Version} \\
        \hline
        Python & 3.10.2 \\
        PyTorch & 2.3.0+cu121 \\
        MONAI & 1.3.0 \\
        SimpleITK & 2.3.1 \\
        PyDICOM & 2.4.4 \\
        SciPy & 1.13.0 \\
        NumPy & 1.26.4 \\
        \hline
    \end{tabular}
    \label{tab:software_env}
    \caption{Core dependencies and library versions.}
\end{table}

\end{document}